\documentclass[a4paper,twocolumn,11pt,unpublished,allowtoday]{quantumarticle}
\pdfoutput=1
\usepackage[utf8]{inputenc}
\usepackage[english]{babel}
\usepackage[T1]{fontenc}
\usepackage{amsmath,amssymb,amsthm,mathtools,booktabs,enumitem,xcolor}
\usepackage{microtype}
\usepackage{graphicx}
\usepackage{caption}
\usepackage{tikz}
\usepackage{quantikz}
\usetikzlibrary{arrows.meta,calc,fit}
\usepackage[numbers,sort&compress]{natbib}
\usepackage{hyperref}
\usepackage[capitalise,noabbrev]{cleveref}
\newtheorem{theorem}{Theorem}
\newtheorem{lemma}[theorem]{Lemma}

\theoremstyle{definition}
\newtheorem{definition}[theorem]{Definition}
\newtheorem{example}[theorem]{Example}
\theoremstyle{remark}

\newcommand{\F}{\mathbb F_2}
\newcommand{\CX}{\mathrm{CX}}
\newcommand{\CZ}{\mathrm{CZ}}
\newcommand{\SWAP}{\mathrm{SWAP}}
\renewcommand{\ket}[1]{\lvert #1\rangle}

\newcommand{\ellis}{\textsc{ELLIS}}
\title{Ellis: Extremely Fast Quantum Circuit Simulation under Qubit Loss and Leakage}
\author{Yuewen Hou}
\email{isaachyw@umich.edu}
\author{Gokul Subramanian Ravi}
\affiliation{University of Michigan, Ann Arbor, USA}

\begin{document}
\maketitle

\begin{abstract}
Qubit loss and leakage are persistent errors: while a qubit remains lost or
leaked, every operation acting on it is dropped.
Consequently, the Clifford circuit executed differs from shot to shot.
This variation breaks Pauli-frame simulation, which constructs each shot by
XORing sampled Pauli flips onto a single \emph{common reference} measurement
record.
By rethinking the roles of tableau simulation and Pauli-frame simulation, we
show that many key fault-tolerant primitives nevertheless retain a common
reference under loss.
These include CSS code memory, lattice surgery, and transversal CNOT.
Such circuits can therefore be simulated exactly under Pauli noise and loss by a
Pauli frame with per-shot masks for the dropped operations.
We define \emph{frame-CSS} circuits as those that are CSS up to local
single-qubit basis changes, given a loss model that specifies where qubits can
be lost and reloaded.
We prove that every frame-CSS circuit has a common reference, and we give a
practically near-linear-time procedure that certifies whether a circuit is frame-CSS.
We implement the certifier and a high-performance simulator in the open-source
package \href{https://github.com/isaachyw/ELLIS}{\textsc{Ellis}} (Efficient
Loss and Leakage Invariant-reference Simulation).
On the gross code under qubit loss and Pauli errors, \ellis{} samples 952{,}000 shots
per second on a single CPU core, at least $1000\times$ faster than
Clifft~\cite{chase2026} and PPVM~\cite{luo2026ppvm}.
\end{abstract}

\section{Introduction}\label{sec:introduction}

Estimating logical failure rates in quantum error correction (QEC) requires a
large number of circuit samples.
Under stochastic Pauli noise, this sampling is efficient.
Every shot executes the same noiseless Clifford circuit.
Stim therefore computes a single noiseless measurement reference by tableau
simulation and represents each shot as a Pauli-frame difference from that
reference~\cite{gidney2021}.

Qubit loss breaks this structure.
Once a qubit is lost, every subsequent operation acting on it is dropped until
the qubit is reloaded.
The executed Clifford circuit therefore depends on the loss pattern of each shot
(\cref{fig:loss-gate-drop}).
Leakage admits a similar model, although a leaked qubit may additionally
propagate errors.
Quasiprobability simulation of non-Pauli noise faces the same difficulty,
because its sampled Clifford errors also alter the circuit from shot to
shot~\cite{hou2026}.
In all three cases, the shots no longer appear to share a reference.
The natural alternative, a separate tableau simulation for each shot, forfeits
the efficiency that makes Pauli-frame sampling practical.

This paper shows that for many important circuits the common reference exists.
Key fault-tolerant primitives, including CSS code memory, lattice surgery, and
transversal CNOT, share a single reference across all loss patterns.
A Pauli frame with per-shot masks for the dropped operations therefore simulates
them exactly.
We identify a structural class of such circuits, the frame-CSS circuits, and
give a procedure that decides membership in this class under stated assumptions.
We implement the certification procedure and a high-performance simulator in the
open-source package \ellis{}, which samples the gross code under loss at 952{,}000 shots per second on a single CPU core.

The rest of the paper is organized as follows.
\Cref{sec:pauli-simulation} describes the distinct information that tableau and
Pauli-frame simulation supply.
\Cref{sec:criterion} explains why loss complicates this division and formalizes the
condition under which a lossy circuit can still be simulated with a masked Pauli
frame.
\Cref{sec:css} shows that being a CSS circuit is sufficient for a common reference.
\Cref{sec:compiler} extends this result to circuits that become CSS after local
single-qubit basis changes.
Gidney and Jones~\cite{gidney2026css} call such circuits \emph{secretly CSS}, a notion they leave
informal and whose detection they leave open.
We formalize it as frame-CSS, which is defined relative to a loss model in our framework because
only operations at which a loss may already have occurred must be CSS, and
reloads constrain the frames where they occur.
When any operation may be followed by a loss and qubits are never reloaded,
frame-CSS corresponds to secretly CSS.
\Cref{sec:compiler} gives a procedure that decides membership under the most
pessimistic loss model, in which a qubit may be lost or reloaded at any point.
Membership there is sufficient for a common reference under every loss model,
but not necessary.
\Cref{sec:evaluation} evaluates \ellis{} against Clifft and PPVM.

\section{Clifford circuit simulation under Pauli errors}\label{sec:pauli-simulation}

We consider Clifford circuits built from $H$, $S$, and $\CX$, with Pauli
preparations and measurements, and with every qubit initialized to $\ket0$.
Each shot produces a record $m\in\F^M$ containing one bit per measurement, where
the bit values $0$ and $1$ denote the eigenvalues $+1$ and $-1$.
We initially omit Pauli gates and Pauli faults, and reintroduce them at the end
of this section.

\subsection{Noiseless records form an affine space}

In a noiseless Clifford circuit, only some measurement outcomes are independent
random bits, and the remaining outcomes are determined by them.
Tableau simulation expresses the record as $a+B\xi$, where $\xi$ is a vector of
independent, uniformly random bits~\cite{aaronson2004}.
The possible noiseless records therefore form an affine space $A=a+V$ with
$V=\operatorname{im}B$, and all records in $A$ are equally likely.

This structure constrains certain parities of the record.
A parity $y\cdot m$ is \emph{deterministic} if it takes the same value on every
possible record, which holds exactly when $y\perp V$.
We collect a basis of these parities as the rows of a \emph{detector matrix} $H$,
and let $\boldsymbol\sigma$ hold the values they take.
Each row of $H$, together with its entry of $\boldsymbol\sigma$, is a
\emph{signed detector}.
Then
\begin{equation}\label{eq:affine}
 \begin{gathered}
 A=\{m:Hm=\boldsymbol\sigma\},\\
 V=\ker H,\qquad V^\perp=\operatorname{rowspan}H.
 \end{gathered}
\end{equation}
Each Clifford circuit thus has its own affine space $A$.
This point is central to what follows: changing or removing a Clifford gate, as
loss does, can change both $H$ and $\boldsymbol\sigma$.

\begin{example}[GHZ records]\label[example]{ex:spaces}
The circuit below prepares a three-qubit GHZ state and measures each qubit in
the $Z$ basis.
\[
\begin{quantikz}[row sep={0.7cm,between origins},column sep=0.35cm]
 \lstick{$\ket0$} & \gate{H} & \ctrl{1} & \ctrl{2} & \meter{} & \rstick{$m_1$}\cw \\
 \lstick{$\ket0$} & \qw & \targ{} & \qw & \meter{} & \rstick{$m_2$}\cw \\
 \lstick{$\ket0$} & \qw & \qw & \targ{} & \meter{} & \rstick{$m_3$}\cw
\end{quantikz}
\]
Its records form the affine space $A=\{000,111\}$, with
\[
 H=\begin{pmatrix}1&1&0\\1&0&1\end{pmatrix},\qquad
 \boldsymbol\sigma=\begin{pmatrix}0\\0\end{pmatrix}.
\]
The record $010$ is impossible, since
$H(010)^{\mathsf T}=(1,0)^{\mathsf T}\neq\boldsymbol\sigma$.
\end{example}

\subsection{Tableau reference and Pauli frame}

Stim samples a circuit under stochastic Pauli noise in two
steps~\cite{gidney2021}.
First, a tableau simulator tracks the full stabilizer state and returns one
noiseless \emph{reference} record $r$.
Because the simulator knows the state, $Hr=\boldsymbol\sigma$.
Second, for every shot, a Pauli-frame simulator tracks only a Pauli operator,
stored as two bits per qubit, that describes how the shot differs from the
reference.
The frame never stores the state.
It can therefore determine which outcomes flip relative to the reference, but
not what the reference outcomes are.
Its noiseless record $v$ is accordingly a difference, satisfying $Hv=0$ rather
than $Hv=\boldsymbol\sigma$.
Attaching a sign to the frame would not help.
Such a sign is a global phase of the difference, whereas $\boldsymbol\sigma$ is
determined by the signs of the state's stabilizers, which only the tableau
tracks.

\begin{example}[Reference and frame]\label[example]{ex:frame}
The circuit below prepares $(\ket{00}-\ket{11})/\sqrt2$ and measures both qubits
in the $X$ basis.
\[
\begin{quantikz}[row sep={0.7cm,between origins},column sep=0.35cm]
 \lstick{$\ket0$} & \gate{H} & \ctrl{1} &          &          & \gate{M_X} & \rstick{$m_0$}\cw \\
 \lstick{$\ket0$} &          & \targ{}  & \gate{S} & \gate{S} & \gate{M_X} & \rstick{$m_1$}\cw
\end{quantikz}
\]
Since $X_0X_1$ stabilizes this state with sign $-1$, the two readouts always
differ.
Hence $A=\{01,10\}$, $H=(1\;1)$, and $\boldsymbol\sigma=1$, and a tableau run may
return $r=01$.

The frame stores the bits $(z,x)$ for each qubit.
Each reset initializes its qubit's frame with a random $Z$, which leaves $\ket0$
unchanged; we denote the two random bits by $b_0$ and $b_1$.
The update rules are as follows: $H$ swaps $x$ and $z$; $\CX_{c\to t}$ sets
$x_t\mathrel{\oplus}=x_c$ and $z_c\mathrel{\oplus}=z_t$; $S$ sets
$z\mathrel{\oplus}=x$; and an $X$ readout is flipped by $z$.
\begin{center}
\small
\begin{tabular}{@{}lcc@{}}
\toprule
Step & qubit 0 $(z,x)$ & qubit 1 $(z,x)$\\
\midrule
reset & $(b_0,0)$ & $(b_1,0)$\\
$H_0$ & $(0,b_0)$ & $(b_1,0)$\\
$\CX_{0\to1}$ & $(b_1,b_0)$ & $(b_1,b_0)$\\
$S_1$ & $(b_1,b_0)$ & $(b_1\oplus b_0,b_0)$\\
$S_1$ & $(b_1,b_0)$ & $(b_1,b_0)$\\
\bottomrule
\end{tabular}
\end{center}
Both readouts are flipped by $b_1$, so $v=(b_1,b_1)$, while $b_0$ flips nothing.
The random bit $b_1$ makes each outcome random, and because it flips both
together, the frame reproduces their correlation.
Its own parity, however, is $Hv=0$ in every shot, never $\boldsymbol\sigma=1$.
That sign comes only from the reference: $m=r\oplus v\in\{01,10\}$, as required,
whereas $r=00$ would yield the invalid records $00$ and $11$.
\end{example}

The same holds for any noiseless Clifford circuit.
The frame maps the reference outcome to another possible outcome, so
$r\oplus v\in A$ and $Hv=H(r\oplus v)\oplus Hr=0$.

Pauli faults, including explicit Pauli gates and readout errors, contribute a
further flip vector $d$.
Each shot is then
\begin{equation}\label{eq:noisy-frame}
 m=r\oplus v\oplus d,\qquad Hm\oplus\boldsymbol\sigma=Hd.
\end{equation}
Detection events therefore depend only on the faults.
A single tableau run serves every shot, and each shot requires only the fast,
linear frame update, XORed onto $r$.

\begin{center}
\fbox{\parbox{\dimexpr\columnwidth-2\fboxsep-2\fboxrule\relax}{%
\textbf{Takeaway.} Under stochastic Pauli noise, once a valid reference $r$ is
known, each shot requires only a Pauli-frame simulation XORed onto $r$; no
further tableau simulation is needed.}}
\end{center}

\paragraph{Cost.}
For a circuit with $N$ qubits, $G$ gates, and $J$ measurements and resets, a
dense tableau stores $O(N^2)$ bits and computes the reference in
$O(N^2+GN+JN^2)$ time~\cite{aaronson2004}.
A Pauli frame stores $O(N)$ bits and simulates each shot in $O(N+G+J)$ time.
Because frames are algebraically simple, the frames of many shots can be
bit-packed and updated together with SIMD instructions~\cite{gidney2021}.

\section{Qubit loss and the common-reference condition}\label{sec:criterion}

This section formalizes the loss and leakage error model and explains why it is
harder to simulate than stochastic Pauli noise.
A qubit is \emph{lost} when it disappears from the device and is not immediately
replaced, and it is \emph{leaked} when it leaves the computational subspace.
Both errors can persist until they are detected and corrected.
They arise in neutral-atom systems, trapped-ion hyperfine qubits, and
superconducting transmons~\cite{baranes2026,hayes2020,varbanov2020}.
How a gate interacts with an unavailable qubit depends on the platform.
We adopt the gate-cancellation model used in atom-loss
studies~\cite{baranes2026}, and our analysis applies equally to leakage whenever
it is represented by this model.
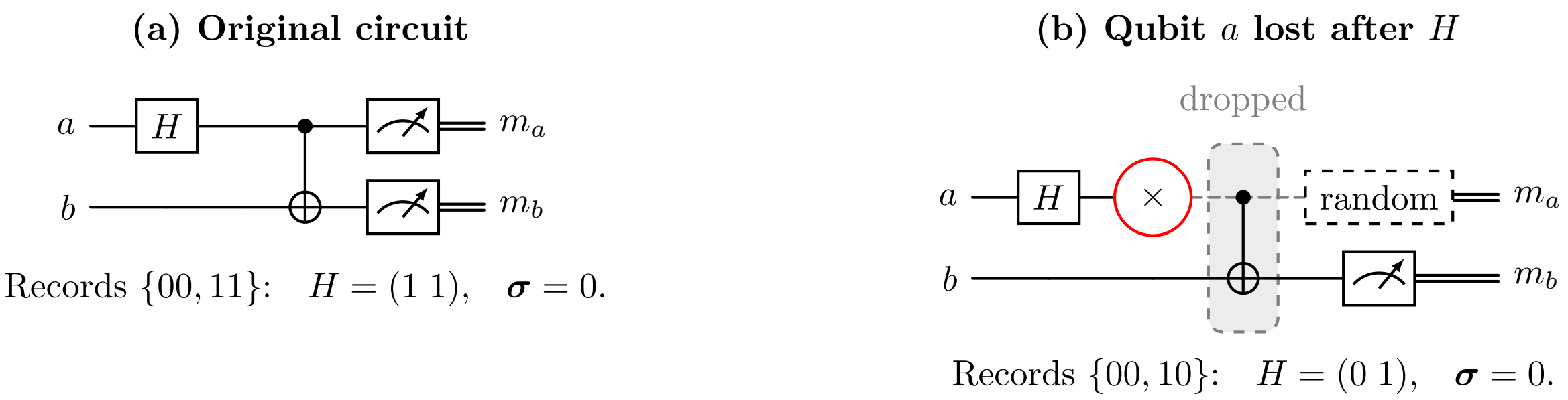
\begin{figure*}[t]
\centering
\newcommand{\lostwire}{\wireoverride{n}\wire[l][1][gray,densely dashed]{q}}
\begin{minipage}[t]{.45\textwidth}
\centering\small
\textbf{(a) Original circuit}\par\medskip
\begin{quantikz}[row sep={0.8cm,between origins},column sep=0.45cm]
 \lstick{$a$} & \gate{H} & & \ctrl{1} & \meter{} & \rstick{$m_a$}\setwiretype{c} \\
 \lstick{$b$} &          & & \targ{}  & \meter{} & \rstick{$m_b$}\setwiretype{c}
\end{quantikz}\par\medskip
Records $\{00,11\}$:\quad $H=(1\;1)$,\quad $\boldsymbol\sigma=0$.
\end{minipage}\hfill
\begin{minipage}[t]{.45\textwidth}
\centering\small
\textbf{(b) Qubit $a$ lost after $H$}\par\medskip
\begin{quantikz}[row sep={0.8cm,between origins},column sep=0.45cm]
 \lstick{$a$} & \gate{H}
   & \gate[style={circle,draw=red,text=red,inner sep=0pt,minimum size=4mm}]{\times}
   & \ctrl{1}\lostwire
     \gategroup[2,steps=1,style={dashed,rounded corners,draw=gray,fill=gray!15,inner xsep=2pt},background,label style={label position=above,anchor=south,yshift=-2pt,text=gray}]{dropped}
   & \gate[style={dashed}]{\text{random}}\lostwire & \rstick{$m_a$}\setwiretype{c} \\
 \lstick{$b$} & & & \targ{} & \meter{} & \rstick{$m_b$}\setwiretype{c}
\end{quantikz}\par\medskip
Records $\{00,10\}$:\quad $H=(0\;1)$,\quad $\boldsymbol\sigma=0$.
\end{minipage}
\caption{Gate cancellation after loss. $\boldsymbol\sigma$ gives the expected parities. (a)~Without loss, the readouts agree, so the fixed parity is $m_a\oplus m_b=0$.
(b)~Qubit $a$ is lost after $H$, so the $\CX$ is omitted.
The lost qubit's record slot receives a random bit, and $b$ stays in $\ket0$, so the fixed parity becomes $m_b=0$.
The loss changes the detector matrix $H$.}
\label{fig:loss-gate-drop}
\end{figure*}

\Cref{fig:loss-gate-drop} shows how a loss changes the circuit.

Unlike a Pauli fault, a dropped Clifford gate changes the ideal circuit itself,
not only its Pauli deviation.
Existing simulators therefore either update their Clifford coordinates dynamically for each
shot~\cite{chase2026} or run a full tableau simulation for each
shot~\cite{luo2026ppvm}.

\begin{definition}[Lossy circuit]\label[definition]{def:lossy}
A \emph{loss pattern} $s$ lists the points after which qubits are lost, and the
reload points at which an empty slot receives a fresh qubit.
The loss model (\cref{sec:loss-model}) specifies which patterns are allowed.
The \emph{lossy circuit} $C_s$ is the original circuit modified as follows.
Losing a qubit discards it, leaving the remaining qubits in the reduced state
given by the partial trace.
While a qubit is lost, every gate and reset acting on it is removed; a two-qubit
gate is removed as a whole, so that its partner is left untouched.
Every readout of a lost qubit records an independent, uniformly random bit in its
own record slot.
A reload places a fresh qubit, in $\ket0$ or in the maximally mixed state, into
an empty slot.
\end{definition}

For each loss pattern $s$, the lossy circuit $C_s$ is still a stabilizer circuit,
so the notation of \cref{sec:pauli-simulation} applies with a subscript $s$:
\begin{equation}\label{eq:lossy-spaces}
 \begin{gathered}
 A_s=a_s+V_s=\{m:H_s m=\boldsymbol\sigma_s\},\\
 V_s=\ker H_s,\qquad V_s^\perp=\operatorname{rowspan}H_s.
 \end{gathered}
\end{equation}
Dropped gates and discarded qubits can change both $H_s$ and
$\boldsymbol\sigma_s$, so a reference that is valid without loss need not lie in
$A_s$.

A \emph{masked} Pauli frame applies each frame update only when the operands of
the operation are present.
It therefore produces the correct variation $v_s\in V_s$ and the correct fault
flips $d_s$ for each shot.
With one shared reference $r$, its records are uniform on $r+d_s+V_s$, whereas
the exact records are uniform on $d_s+A_s$.
Pauli faults and ideal Pauli gates only flip record bits, and the masked frame
tracks them exactly as part of $d_s$.
We therefore omit them from the remaining analysis and consider only the
Pauli-free part of the circuit.
The masked frame is then exact when $r+V_s=A_s$, that is, when
$r\in A_s$, and the only remaining question is whether a single $r$ satisfies
this for every loss pattern.

\begin{definition}[Common reference]\label[definition]{def:common}
A \emph{common reference} is a record $r$ with $r\in A_s$ for every allowed loss
pattern $s$.
\end{definition}

\begin{theorem}[Exact signed-detector criterion]\label{thm:criterion}
A shared reference $r$ makes the masked sampler exact for every allowed loss
pattern and Pauli fault history if and only if
\begin{equation}\label{eq:system}
 \boxed{\quad H_s r=\boldsymbol\sigma_s\quad\text{for every allowed }s.\quad}
\end{equation}
Consequently, no common reference exists if and only if the signed detector
equations $(y_i,\sigma_i)$ of some loss patterns XOR to a contradiction:
\begin{equation}\label{eq:witness}
 \bigoplus_{i=1}^k y_i=0,\qquad \bigoplus_{i=1}^k\sigma_i=1.
\end{equation}
Such a contradiction requires at most $M+1$ equations.
\end{theorem}

Checking \cref{eq:system} directly requires $H_s$ and $\boldsymbol\sigma_s$ for
every loss pattern, and the number of loss patterns grows exponentially with the
number of operations.
CSS circuits avoid this search: as we show next, they always satisfy
\cref{eq:system}.

\section{CSS circuits have a common reference}\label{sec:css}

This section shows that if the Pauli-free part of a circuit is CSS, then the
all-zero record is a common reference under every loss pattern.

A \emph{CSS circuit} prepares qubits in $\ket0$ or $\ket+$, applies positive CSS
Cliffords such as $\CX$ and $\SWAP$ (\cref{lem:css-operations}), and measures in
the $X$ or $Z$ basis.
Under loss, gates may be dropped, qubits discarded, and fresh qubits reloaded in
$\ket0$ or in the maximally mixed state.

\paragraph{The invariant.}
The proof tracks a single property of the quantum state.

\begin{definition}[Positive CSS state]\label[definition]{def:positive-css}
A stabilizer state, possibly mixed (with fewer generators than qubits, as after a
loss), is \emph{positive CSS} if its stabilizer group is generated by operators
of the form $+X^a$ or $+Z^b$.
Here $X^a$ applies $X$ to every qubit in the support of the bit vector $a$, and
$Z^b$ is defined similarly.
We call such generators \emph{pure positive}.
\end{definition}

For example, a Bell pair stabilized by $+XX$ and $+ZZ$ is positive CSS, whereas
one stabilized by $-ZZ$ is not.
The sign condition applies only to the generators; products such as
$(+XX)(+ZZ)=-YY$ may be negative.

\begin{lemma}[CSS operations preserve the positive CSS state]\label[lemma]{lem:css-operations}
Starting from a positive CSS state, each of the following operations yields a
positive CSS state: preparing $\ket0$ or $\ket+$; adding a maximally mixed qubit;
discarding a qubit; and applying a \emph{positive CSS Clifford}, which maps every
$X_i$ to a $+$pure-$X$ operator and every $Z_i$ to a $+$pure-$Z$ operator (for
example, $\CX$ and $\SWAP$).
For an $X$ or $Z$ measurement, outcome $0$ always has nonzero probability, and
the post-measurement state for outcome $0$ is again positive CSS.
\end{lemma}
\begin{proof}
Every element of the stabilizer group can be written as $+X^aZ^b$, where $a$ and
$b$ lie in the spans of the $X$ and $Z$ generator supports.
The $X$ and $Z$ parts commute, so reordering a product of generators introduces
no sign.
In particular, every pure-$X$ or pure-$Z$ stabilizer has sign $+$.
\begin{itemize}[nosep,leftmargin=*]
 \item \textbf{Preparation.} Preparing a fresh qubit adds $+Z_q$ or $+X_q$, and
 adding a maximally mixed qubit adds no generator. Resetting a qubit in use is
 equivalent to discarding it and then preparing it.
 \item \textbf{Discarding.} Discarding qubit $q$ keeps exactly the stabilizers
 that act trivially on $q$. Since the group is $\{X^aZ^b\}$ with $a$ and $b$
 ranging over the spans of the $X$ and $Z$ generator supports, these stabilizers
 form $\{X^a:a_q=0\}\times\{Z^b:b_q=0\}$. This group is again generated by pure
 positive elements.
 \item \textbf{Positive CSS Clifford.} Such a gate maps each $+X^a$ to a product
 of commuting $+$pure-$X$ operators, which is again $+$pure-$X$, and likewise
 maps each $+Z^b$ to a $+$pure-$Z$ operator.
 \item \textbf{$Z_q$ measurement.} If $\pm Z_q$ is a stabilizer, it must be
 $+Z_q$, so the outcome is $0$. Otherwise, outcome $0$ occurs with probability
 $1/2$. If some generators anticommute with $Z_q$, they are all $X$-type. We
 choose one of them as the pivot and multiply each of the others by it, which
 keeps them positive and pure; the update then replaces the pivot with $+Z_q$.
 If no generator anticommutes with $Z_q$, the update simply adds $+Z_q$. An
 $X_q$ measurement is handled identically, with $X$ and $Z$
 exchanged.\qedhere
\end{itemize}
\end{proof}

\begin{theorem}[Zero is a common reference for CSS circuits]\label{lem:css}
If the Pauli-free part of a circuit is CSS and starts in a positive CSS state,
then for every loss and reload pattern $s$,
\[
 0\in A_s\qquad\text{and}\qquad \boldsymbol\sigma_s=0.
\]
\end{theorem}
\begin{proof}
Run $C_s$, choosing outcome $0$ at every random measurement and at every readout
of a lost qubit.
Dropped gates have no effect, and every other step is covered by
\cref{lem:css-operations}.
The state therefore remains positive CSS, and every chosen outcome has nonzero
probability.
Deterministic measurements on this branch also yield $0$.
Hence the all-zero record is possible, so $0\in A_s$ and
$\boldsymbol\sigma_s=H_s0=0$.
\end{proof}

\Cref{fig:loss-gate-drop} is an instance of this theorem, since its $H$ on
$\ket0$ simply prepares $\ket+$.
Losing qubit $a$ changes the detector matrix from $(1\;1)$ to $(0\;1)$, but
$\boldsymbol\sigma=0$ in both cases, so $r=0$ is valid for both.

Consequently, a CSS circuit requires no tableau simulation at all.
With $r=0$, each shot is simply the output of the masked frame,
\[
 m=v_s\oplus d_s,\qquad H_s m=H_s d_s,
\]
which follows the exact noisy distribution of $C_s$.

\paragraph{Circuits compiled to other gate sets.}
Many circuits with CSS structure are nevertheless compiled with other gates:
$\CZ$ in place of $\CX$, Hadamards around a measurement, or transversal logical
gates that rotate the Pauli basis of each qubit.
Such gates break the invariant in the original basis.
For example, a Hadamard on one half of the Bell pair above maps its stabilizers
$+XX$ and $+ZZ$ to $+XZ$ and $+ZX$, which are not pure.
Often, however, the circuit becomes CSS after a suitable single-qubit change of
basis on each qubit.
\Cref{sec:compiler} formalizes these \emph{frame-CSS} circuits and shows how to
find such basis changes or prove that none exist.

\section{Frame-CSS circuits}\label{sec:compiler}

\begin{figure*}[t]
\centering
\begin{minipage}[t]{.48\textwidth}
\centering\small
\textbf{(a) Physical circuit fragment}\par\smallskip
\begin{quantikz}[row sep={0.85cm,between origins},column sep=0.4cm]
 \lstick{$q_a$} & \gate{H} & \ctrl{1} & \gate{M_Z} & \rstick{$m_a$}\cw \\
 \lstick{$q_b$} & \gate{H} & \targ{}  & \gate{M_Z} & \rstick{$m_b$}\cw
\end{quantikz}
\end{minipage}\hfill
\begin{minipage}[t]{.48\textwidth}
\centering\small
\textbf{(b) Local-frame description}\par\smallskip
\begin{quantikz}[row sep={0.85cm,between origins},column sep=0.4cm]
 \lstick{$q_a$} & \gate{I} & \targ{}   & \gate{M_X} & \rstick{$m_a$}\cw \\
 \lstick{$q_b$} & \gate{I} & \ctrl{-1} & \gate{M_X} & \rstick{$m_b$}\cw
\end{quantikz}
\end{minipage}
\caption{Absorbing a transversal logical $H$ into frames. Two data wires of a CSS block with
transversal logical $H$ are followed by a CNOT and readouts. Frames $F_q=I$ before the $H$ layer and
$F_q=H$ after it absorb the $H$ gates, since $(H\otimes H)\CX_{a\to b}(H\otimes H)=\CX_{b\to a}$ and $HZH=X$.
The framed circuit (b) is positive CSS with $\kappa=0$. The identity placeholders keep the physical time
steps, so a loss between $H$ and readout is still modeled.}
\label{fig:logical-h-frames}
\end{figure*}
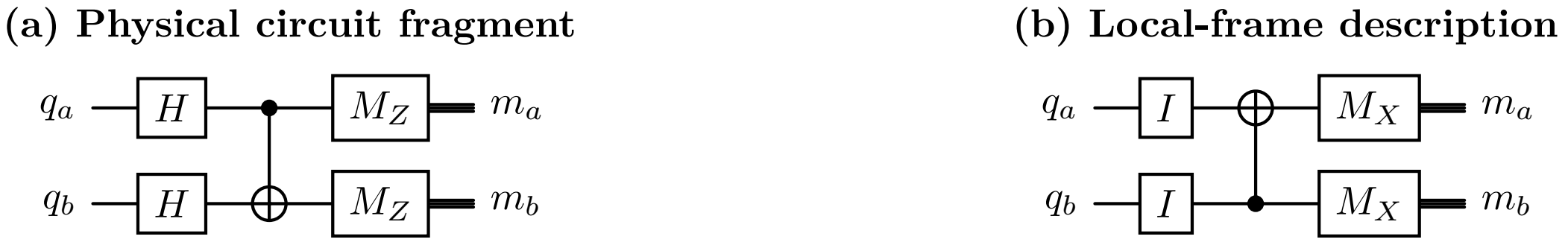

Many practical circuits are not CSS as written: XZZX surface codes mix $X$ and $Z$ checks, transversal logical gates rotate each qubit's Pauli basis, and hardware often provides $\CZ$ rather than $\CX$.
Such circuits often become CSS after a fixed single-qubit change of basis on each qubit.
In \cref{fig:logical-h-frames}, relabeling both qubits by $H$ after the Hadamard layer turns the physical $\CX$ and $Z$ readouts into a reversed $\CX$ and $X$ readouts.
The relabeling changes only the description, not the physical schedule, so losses can occur at the same places.
We call such circuits \emph{frame-CSS}, prove that they have a common reference, and show how \textsc{Ellis} recognizes them.

\subsection{Frames}

A \emph{frame} is a single-qubit Clifford $F$ with $\ket{\psi}=F\ket{\psi_{\mathrm{loc}}}$: the qubit reads the physical Paulis $FXF^\dagger$ and $FZF^\dagger$ as its local $X$ and $Z$.
A single-qubit gate $g$ changes the frame to $gF$ and disappears from the local description; in \emph{reset, $H$, measure $Z$} with initial frame $I$, the readout measures $H^\dagger ZH=X$ in the frame.
A two-qubit gate leaves frames unchanged, so its local form depends on the frames of both operands: $\CZ_{ab}$ is not CSS, but in frames $I$ on $a$ and $H$ on $b$ it is $(I\otimes H)\,\CZ_{ab}\,(I\otimes H)=\CX_{a\to b}$.

The operations before the first possible loss form the loss-free \emph{prefix}, such as a noiseless encoder; no qubit is lost there, so only the state it leaves matters (\cref{sec:loss-model}).
Each qubit's timeline is cut into \emph{segments} at every reset, every readout, and the end of the prefix.
Each segment's starting frame is fixed in advance and then follows the segment's single-qubit gates, also while the qubit is lost (\cref{fig:segments}).
Losses and reloads do not cut segments, since frames cannot depend on the shot; a reloaded qubit enters in the frame of that moment.

\DeclareRobustCommand*\condmark[1]{\tikz[baseline=(rm.base)]{\node[circle,draw,inner sep=0.6pt,font=\tiny](rm){#1};}}
\begin{figure}[t]
\centering
\tikzset{seg/.style={dashed,rounded corners,draw=gray,inner xsep=1pt,inner ysep=1pt},
         segl/.style={label position=above,anchor=south,yshift=-1pt,font=\scriptsize},
         fr/.style={draw=none,fill=none,font=\scriptsize,inner sep=0pt},
         rl/.style={draw,circle,fill=none,font=\tiny,inner sep=0.6pt,minimum size=0pt,minimum width=0pt,minimum height=0pt}}
\resizebox{\columnwidth}{!}{%
\begin{quantikz}[row sep={0.95cm,between origins},column sep=0.22cm]
 \lstick{$a$} & \gate{R}\gategroup[1,steps=4,style=seg,label style=segl]{choose $F_1$}
   & \gate{H} & \ctrl{1} & \gate{M_Z}
   & \gate{R}\gategroup[1,steps=4,style=seg,label style=segl]{choose $F_2$}
   & \gate{H} & \ctrl{1} & \gate{M_Z} \\
 \lstick{$d$} & \gate{R} & & \targ{} & & & & \targ{} & \\
 \lstick{frame of $a$}\setwiretype{n}
   & \gate[style=fr]{F_1} & \gate[style=fr]{HF_1} & \gate[style=fr]{HF_1} & \gate[style=fr]{HF_1}
   & \gate[style=fr]{F_2} & \gate[style=fr]{HF_2} & \gate[style=fr]{HF_2} & \gate[style=fr]{HF_2} \\
 \lstick{frame of $d$}\setwiretype{n}
   & \gate[style=fr]{G} & \gate[style=fr]{G} & \gate[style=fr]{G} & \gate[style=fr]{G}
   & \gate[style=fr]{G} & \gate[style=fr]{G} & \gate[style=fr]{G} & \gate[style=fr]{G} \\
 \lstick{condition}\setwiretype{n}
   & \gate[style=rl]{b} & & \gate[style=rl]{a} & \gate[style=rl]{c}
   & \gate[style=rl]{b} & & \gate[style=rl]{a} & \gate[style=rl]{c}
\end{quantikz}%
}
\caption{Segments and frames. Ancilla $a$ is reset and measured twice, and data qubit $d$ is reset once.
Each timeline is cut at every reset and every readout. The segments that begin at a readout of $a$ hold no gates here, so only the two that begin at a reset are drawn, with starting frames $F_1$ and $F_2$; $d$ has one segment, with frame $G$.
The frame rows give each qubit's frame after every operation: the $H$ gate carries $F_i$ to $HF_i$, and the frame of $d$ stays $G$ because $d$ has no single-qubit gate.
Circled letters mark where conditions (a)--(c) of \cref{def:frame-css} constrain the three frames.
\protect\condmark{a}~Each $\CX$ must be a positive CSS Clifford in the frames, so $HF_i$ and $G$ must read the control's $Z$ and the target's $X$ as different local Paulis.
\protect\condmark{b}~Each reset must prepare $+X$ or $+Z$ in its frame: $F_1$ and $F_2$ for $a$, and $G$ for $d$.
\protect\condmark{c}~Each readout of $a$ must measure $\pm X$ or $\pm Z$ in frame $HF_i$, and its sign fixes the reference bit $\kappa_m$. The empty segment after the readout must keep that sign, but the reset that follows discards the state, so only \protect\condmark{b} constrains $F_2$.}
\label{fig:segments}
\end{figure}
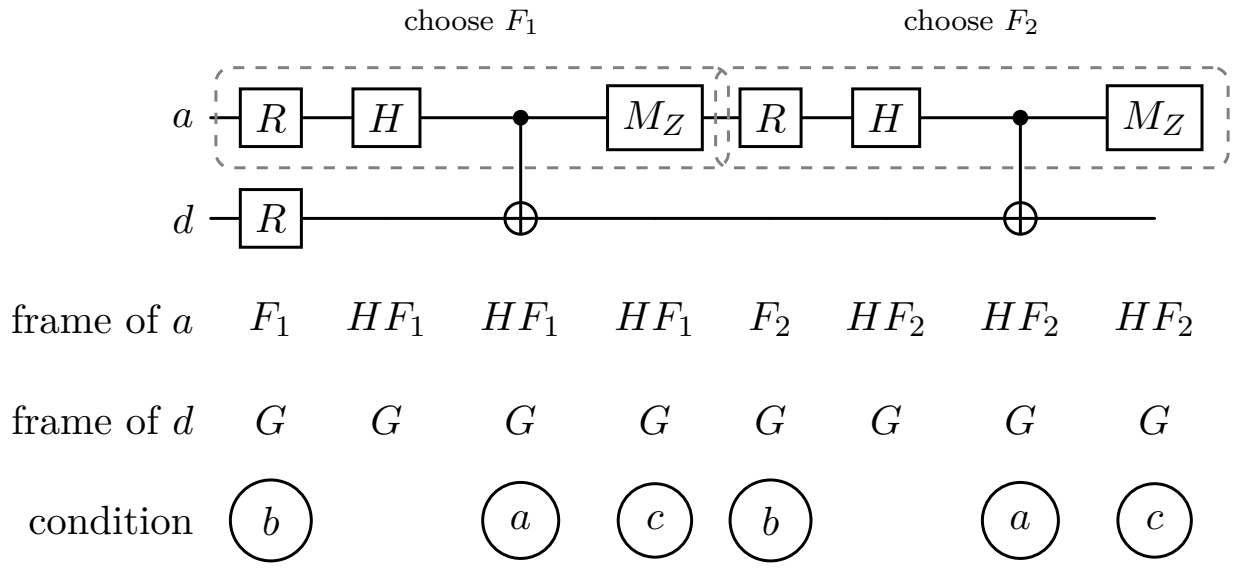

\begin{definition}[Frame-CSS]\label[definition]{def:frame-css}
A circuit is \emph{frame-CSS} for a loss model if one frame per segment can be chosen such that, in the frames:
\begin{enumerate}[label=(\alph*),leftmargin=*,nosep]
 \item every two-qubit gate is a positive CSS Clifford (\cref{lem:css-operations});
 \item every reset, and every reload into $\ket0$ wherever the loss model allows, prepares $+X$ or $+Z$;
 \item every readout measures $(-1)^{c_m}Q$ for some $Q\in\{X,Z\}$ and sign bit $c_m\in\{0,1\}$, giving the reference bit $\kappa_m=c_m$; if the readout does not also reset the qubit, the next segment's frame reads the measured Pauli with the same sign;
 \item the prefix leaves a positive CSS state on the qubits it uses again.
\end{enumerate}
\end{definition}

A qubit is used again if its segment after the prefix contains a two-qubit gate or a readout, and readouts inside the prefix take their bits from one loss-free record.
A reload into the maximally mixed state needs no condition, since it fixes no stabilizer sign.

\begin{theorem}[Common reference from frames]\label{thm:compiler}
If a circuit is frame-CSS, its reference bits $\kappa$ form a common reference: $H_s\kappa=\boldsymbol\sigma_s$ for every allowed loss and reload pattern $s$.
\end{theorem}
\begin{proof}
Let $s$ be an arbitrary allowed loss and reload pattern; we show $\kappa\in A_s$, that is, $C_s$ outputs the record $\kappa$ with nonzero probability.
Describe $C_s$ in the frames and follow the branch on which every readout outputs its bit of $\kappa$.
We show that along this branch the local state stays positive CSS and each readout outputs its bit with nonzero probability given the earlier ones; multiplying these probabilities gives $\kappa\in A_s$.

\emph{Prefix.} The prefix has no losses, so it runs as in the loss-free circuit, and its readouts output their bits of $\kappa$, which come from a loss-free record.
By (d), the state it leaves is positive CSS once the qubits it does not use again are discarded (\cref{lem:css-operations}, discarding).

\emph{After the prefix}, each operation of $C_s$ is one of the following.
\begin{itemize}[leftmargin=*,nosep]
 \item \emph{Single-qubit gate:} it only changes the frame, so the local state is unchanged.
 \item \emph{Two-qubit gate:} if both operands are present, it is a positive CSS Clifford by (a) (\cref{lem:css-operations}, Clifford); if an operand is lost, $C_s$ drops the gate.
 \item \emph{Loss:} the qubit is discarded (\cref{lem:css-operations}, discarding).
 \item \emph{Reset, or reload into $\ket0$:} by (b), it prepares $+X$ or $+Z$, that is, $\ket+$ or $\ket0$ in the frame, after discarding any old state (\cref{lem:css-operations}, discarding and preparing).
 \item \emph{Reload into the maximally mixed state:} \cref{lem:css-operations}, adding a maximally mixed qubit.
 \item \emph{Readout of a present qubit:} by (c), it measures $(-1)^{c_m}Q$, so outputting $\kappa_m$ is outcome $0$ of a local $Q$ measurement. That outcome is possible and leaves a positive CSS state (\cref{lem:css-operations}, measurement).
 \item \emph{Readout of a lost qubit:} it outputs a fair coin, which equals $\kappa_m$ with probability $1/2$; the state is unchanged.
 \item \emph{Frame change after a readout without reset:} the measured qubit, if present, is alone in the $+Q$ eigenstate, and by (c) the next segment's frame reads it as $+X$ or $+Z$, so the change discards the qubit and prepares that state (\cref{lem:css-operations}, discarding and preparing). After a readout that also resets, the reset case applies.
\end{itemize}
\end{proof}

\Cref{lem:css} is the special case in which every frame is the identity.

\subsection{How \textsc{Ellis} finds frames}

Up to a Pauli, a frame is fixed by the physical Paulis it reads as local $X$ and $Z$, so there are six.
\textsc{Ellis} narrows the frames in four stages; each keeps the frames that (a)--(c) allow, or finds a conflict proving that none exist.
\begin{enumerate}[label=\arabic*.,leftmargin=*,nosep]
 \item \textbf{Classify gates.} Each two-qubit gate type is checked once against all $36$ pairs of frames.
 A \emph{controlled} gate ($\CX$, $\CZ$, CY, and variants) acts as a $P$-controlled $Q$ and is CSS when the frames read $P$ on the control and $Q$ on the target as different local Paulis.
 A \emph{permutation} gate ($\SWAP$, $\CX\SWAP$, $\SWAP\CX$) is CSS when one operand's frame is a fixed Clifford times the other's.
 Any other gate, such as iSWAP or $\sqrt{XX}$, is CSS in no frames and rejects the circuit.
 \item \textbf{Group segments.} A weighted union--find merges the segments tied by permutation gates and records the Clifford relating their frames.
 A loop whose Cliffords do not compose to a Pauli rejects the circuit, as does a group whose operations involve all three physical Paulis.
 \item \textbf{Orient groups.} A group's \emph{orientation} says which of its at most two physical Paulis reads as local $X$.
 Controlled gates fix relative orientations, so orienting is a 2-coloring problem, solved by a parity union--find; an odd cycle rejects the circuit.
 Each connected \emph{component} is oriented up to swapping $X$ and $Z$ on the whole component.
 \item \textbf{Fix signs.} What remains is a Pauli per group, which only flips signs.
 Each sign requirement in (a)--(d) is an affine equation over $\mathbb F_2$, and Gaussian elimination either solves the system, giving $\kappa$, or proves that no frames exist.
\end{enumerate}

\paragraph{Assumptions.}
\textsc{Ellis} takes circuits of single-qubit Pauli-basis resets and readouts, single- and two-qubit Clifford gates, and Pauli noise, with no classically controlled gates.
It checks frame-CSS under the most pessimistic loss model only: a qubit may be lost before or after any operation, and a fresh qubit in $\ket0$ may be reloaded between any two operations.
Every pattern of another loss model is one of these or equivalent to one (a maximally mixed reload is a $\ket0$ reload followed by a random $X$), so by \cref{thm:compiler} an accepted circuit has a common reference under every loss model.

\paragraph{Cost.}
With union by size and path compression, Stages~1--3 take $O(T\alpha(T))$ time for $T$ operations.
Stage~4 uses Gaussian elimination, which is fast in practice but quadratic in the number of groups in the worst case.
On our largest benchmark, the 932-qubit chgp\_15 memory circuit with $2\times10^4$ operations, the whole check takes $1.796$\,s.

\subsection{Dependence on the loss model}\label{sec:loss-model}

A weaker loss model imposes fewer conditions, so the pessimistic check is sufficient but not necessary.
Condition (b) must hold wherever a reload may occur, not only at segment boundaries.
For example, $HSSH\cdot HSSH=I$ leaves the frame unchanged, but between the two halves the frame is $HSSH=X$, so a qubit reloaded there enters as $\ket1$ in the frame and violates (b).
\ellis{} therefore checks (b) after every single-qubit gate on a qubit that may be missing.
Without reloads, the same circuit is frame-CSS.
Likewise, a model with a loss-free prefix, such as a noiseless encoder, requires only (d) of the prefix, not CSS gates.

\section{Evaluation}\label{sec:evaluation}

We evaluate \ellis{} on two questions: how many protocols the certifier admits,
and how quickly certified circuits can be sampled.
Unless stated otherwise, simulations run on Perlmutter CPU nodes, each with two
AMD EPYC 7763 processors and 128 physical cores.
We use either one core or all 128, with one sampling thread per core.
Our experiments model qubit loss only.
Under the gate-cancellation model, a leaked qubit is handled identically, so the
same certification and sampling apply to leakage without modification.

\subsection{Certification}

We run the certifier on 73 circuits (\cref{tab:lightstim-categories}): a
distance-three export of each of the 43 LightStim~\cite{fang2026lightstim}
protocols, and 30 two-round qLDPC memory circuits generated with
qLDPC~\cite{perlin2023qldpc}.
The frame-CSS check uses the pessimistic model of \cref{sec:compiler}.
The contradiction search uses a weaker model, in which any qubit that an
operation acts on may be lost after each gate, reset, or measurement, and lost
qubits are not reloaded; a contradiction there also excludes a common reference
under the pessimistic model.

In total, the certifier establishes a common reference for 62 of the 73
circuits, proves that none exists for nine, and leaves two open.
All 62 are frame-CSS, although only three are CSS in their native basis.
The frame-CSS circuits include every qLDPC memory built on a CSS code, as well as
the XZZX surface code.
The first open circuit, \path{L:logical_s_to_the_fourth_is_identity}, is not
frame-CSS, but its closed $S$/$\CZ$ blocks give it a common reference when qubits
are not reloaded, so it shows that the pessimistic check is not necessary.
The nine circuits without a common reference are Y-state injection, six
$S$-gate teleportation circuits, the $S$ round trip, and the five-qubit code.
For each, we find loss patterns whose signed detector equations XOR to a
contradiction, and we confirm the contradiction with Stim.
The second open case is lattice-surgery distillation, which is not frame-CSS and whose
contradiction search times out after two minutes.
The structural checks finish in under 2\,s on every circuit, and the
contradiction search takes at most 91\,s on the nine circuits without a common
reference.

\onecolumngrid
\begin{center}
\includegraphics[width=0.49\textwidth]{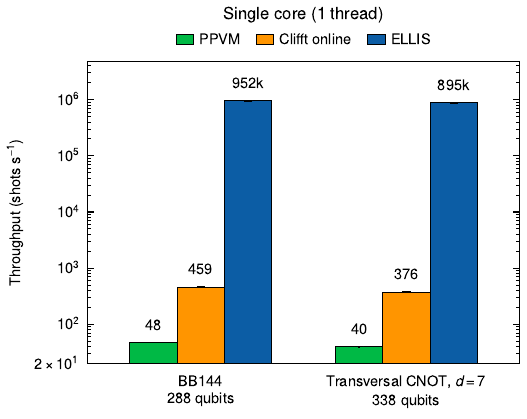}\hfill
\includegraphics[width=0.49\textwidth]{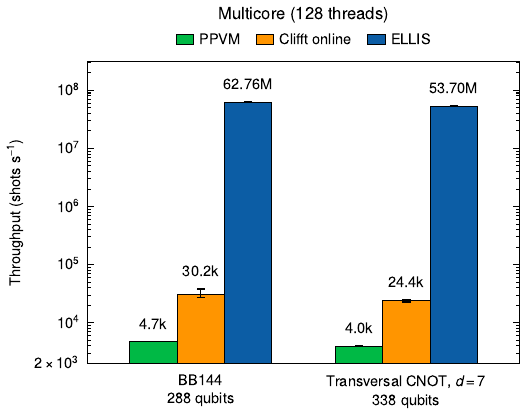}
\captionof{figure}{Sampling throughput on BB144 memory and the distance-7 surface-code
transversal CNOT, using one core (left) and 128 cores (right). Bars show the
median of five repeated batches, and error bars span the observed minimum and
maximum.}
\label{fig:sampling-throughput}
\end{center}
\twocolumngrid

\subsection{Sampling speed on certified circuits}

We compare \ellis{} with PPVM~\cite{luo2026ppvm} and Clifft~\cite{chase2026} on
BB144 memory and the distance-7 surface-code transversal CNOT, using Clifft
0.13.0 in online mode.
The error model adds a loss probability of $10^{-3}$ per two-qubit-gate operand
and circuit-level depolarizing noise at $10^{-3}$.
Reload at reset is disabled.
We measure sampling throughput with packed outputs.
Timing includes output generation but excludes circuit parsing, sampler setup.

The three simulators handle loss in different ways.
PPVM's generalized-tableau backend maintains a separate tableau and per-qubit
loss flags for each shot, and skips operations on lost qubits.
Clifft's online mode updates its Clifford coordinates during sampling to account
for dropped gates.
In \ellis{}, a certified common reference allows the masked Pauli-frame updates
of many shots to be batched together.
On one core, \ellis{} is therefore about three orders of magnitude faster than
Clifft and four orders faster than PPVM on these CSS circuits
(\cref{fig:sampling-throughput}).
On BB144, \ellis{} samples about 952{,}000 shots per second on a single core.
On 128 cores, the advantage remains three to four orders of magnitude.

\section{Conclusion}

This paper formalized the conditions under which a Clifford circuit subject to
qubit loss or leakage can still be simulated with a masked Pauli frame.
We gave a procedure that decides, under stated assumptions, whether a circuit
satisfies these conditions, and we released the open-source simulator
\href{https://github.com/isaachyw/ELLIS}{\textsc{Ellis}}.
Beyond simulation, the invariant underlying frame-CSS circuits may also benefit
decoding, because the detector error model likewise depends on the loss
pattern.
We leave this direction to future work.

During the preparation of this manuscript, we became aware of concurrent
work~\cite{gidney2026css} that simulates CSS circuits at $O(1)$ cost per gate,
in a manner closely related to Pauli-frame simulation.
Our results generalize the conditions under which this approach applies.
We also formalize their secretly CSS circuits as frame-CSS circuits, relative to
a loss model, and give a procedure that identifies them, which that work leaves
open.

\section*{Acknowledgments}

This material is based upon work supported by the U.S. Department of Energy, Office of Science, Office of Advanced Scientific Computing Research, Accelerated Research in Quantum Computing under Award Number DE-SC0025633. 
This research was, in part, supported by the DARPA HARQ Program, under Agreement No. HR0011-26-9-E096. 
This material is based upon work supported by the U.S. National Science Foundation under Cooperative Agreement No. 2547483 through a subaward to the University of Michigan.
This research used resources of the National Energy Research Scientific Computing Center, a DOE Office of Science User Facility supported by the Office of Science of the U.S. Department of Energy under Contract No. DE-AC02-05CH11231 using NERSC award NERSC DDR-ERCAP0035341. This work was supported in part by Advanced Micro Devices, Inc. under the AMD University Program’s AI \& HPC Cluster. This research was, in part, funded by the U.S. Government. The views and conclusions contained in this document are those of the authors and should not be interpreted as representing the official policies, either expressed or implied, of the U.S. Government.

During the preparation of this work, the authors used Claude and Codex for assistance with manuscript and code development and proof verification. The authors reviewed and verified all results and writing, and take full responsibility for the final manuscript.

\clearpage
\bibliographystyle{quantum}
\bibliography{refs}

@article{baranes2026,
  author        = {Baranes, G. and others},
  title         = {Leveraging Qubit Loss Detection in Fault Tolerant Quantum Algorithms},
  journal       = {Phys. Rev. X},
  volume        = {16},
  pages         = {011002},
  year          = {2026},
  eprint        = {2502.20558v3},
  archivePrefix = {arXiv},
  note          = {Fig.~2(b)},
}

@article{aaronson2004,
  author        = {Aaronson, Scott and Gottesman, Daniel},
  title         = {Improved simulation of stabilizer circuits},
  journal       = {Phys. Rev. A},
  volume        = {70},
  pages         = {052328},
  year          = {2004},
  eprint        = {quant-ph/0406196},
  archivePrefix = {arXiv},
}

@article{gidney2021,
  author        = {Gidney, Craig},
  title         = {Stim: a fast stabilizer circuit simulator},
  journal       = {Quantum},
  volume        = {5},
  pages         = {497},
  year          = {2021},
  doi           = {10.22331/q-2021-07-06-497},
}

@misc{gidney2026css,
  author        = {Gidney, Craig and Jones, Cody},
  title         = {{CSS} Circuits: useful quantum circuits that can be simulated with {$O(1)$} cost per gate},
  year          = {2026},
  eprint        = {2611.14944},
  archivePrefix = {arXiv},
}

@misc{hou2026,
  author        = {Hou, Yuewen and Ravi, Gokul Subramanian},
  title         = {Scalable Exact Non-{Pauli} Noise Simulation for Fault-Tolerant Quantum Circuits},
  year          = {2026},
  eprint        = {2610.15975},
  archivePrefix = {arXiv},
}

@misc{chase2026,
  author        = {Chase, B. A. and Labib, F.},
  title         = {Clifft: Fast Exact Simulation of Near-{Clifford} Quantum Circuits},
  year          = {2026},
  eprint        = {2604.27058},
  archivePrefix = {arXiv},
}

@misc{luo2026ppvm,
  author        = {Luo, X.-Z. and others},
  title         = {{PPVM}: A Circuit Simulation Framework for Digital Twins of Quantum Processors},
  year          = {2026},
  eprint        = {2610.18225},
  archivePrefix = {arXiv},
}

@article{hayes2020,
  author        = {Hayes, D. and others},
  title         = {Eliminating Leakage Errors in Hyperfine Qubits},
  journal       = {Phys. Rev. Lett.},
  volume        = {124},
  pages         = {170501},
  year          = {2020},
  eprint        = {1912.13131},
  archivePrefix = {arXiv},
}

@article{varbanov2020,
  author        = {Varbanov, B. M. and others},
  title         = {Leakage detection for a transmon-based surface code},
  journal       = {npj Quantum Inf.},
  volume        = {6},
  pages         = {102},
  year          = {2020},
  eprint        = {2002.07119},
  archivePrefix = {arXiv},
}

@misc{fang2026lightstim,
  author        = {Fang, X. and Wang, M. and Wu, Y. and Prabhu, S. and Tullsen, D. and Miniskar, N. R. and Mueller, F. and Humble, T. and Ding, Y.},
  title         = {{LightStim}: A Framework for {QEC} Protocol Evaluation and Prototyping with Automated {DEM} Construction},
  year          = {2026},
  eprint        = {2604.21472},
  archivePrefix = {arXiv},
}

@misc{perlin2023qldpc,
  author        = {Perlin, M. A.},
  title         = {{qLDPC}: A {Python} toolkit for quantum low-density parity-check codes},
  year          = {2023},
  howpublished  = {\url{https://github.com/qLDPCOrg/qLDPC}},
}
\clearpage
\onecolumn
\appendix
\section{Common-reference certification benchmark}\label{app:lightstim}

\makeatletter
\g@addto@macro\UrlBreaks{\do\_}
\makeatother

\begin{table}[!htbp]
\centering
\footnotesize
\setlength{\tabcolsep}{3pt}
\caption{Common-reference certification.
A check mark indicates a circuit that is frame-CSS under the pessimistic loss model.
Check times are wall seconds.
LightStim~\cite{fang2026lightstim} prefixes abbreviate \texttt{TestMemory\_test\_} (M),
\texttt{TestLogicalOps\_test\_} (L), and \texttt{TestLogicalCircuits\_test\_} (C).}
\label{tab:lightstim-categories}
\begin{tabular}{@{}r p{.27\textwidth} c r @{\hspace{1.5em}} r p{.27\textwidth} c r@{}}
\toprule
No. & Circuit & Frame-CSS & Check & No. & Circuit & Frame-CSS & Check\\
\midrule
1 & \path{M:repetition_code} & \checkmark & 0.017 & 23 & \path{L:two_patch_ls_zz} & \checkmark & 0.077 \\
2 & \path{M:rotated_surface_code_X} & \checkmark & 0.019 & 24 & \path{L:two_patch_ls_xx} & \checkmark & 0.076 \\
3 & \path{M:rotated_surface_code_Z} & \checkmark & 0.016 & 25 & \path{L:multi_patch_ls} & \checkmark & 0.191 \\
4 & \path{M:rotated_surface_code_with_data_defect} & \checkmark & 0.030 & 26 & \path{L:cross_ls} & \checkmark & 0.050 \\
5 & \path{M:unrotated_surface_code_X} & \checkmark & 0.025 & 27 & \path{L:cnot_ls} & \checkmark & 0.145 \\
6 & \path{M:unrotated_surface_code_Z} & \checkmark & 0.024 & 28 & \path{C:bell_teleport_tg_Z} & \checkmark & 0.099 \\
7 & \path{M:toric_code} & \checkmark & 0.039 & 29 & \path{C:bell_teleport_zz_ls_Z} & \checkmark & 0.146 \\
8 & \path{M:color_code} & \checkmark & 0.024 & 30 & \path{C:bell_teleport_xx_ls_X} & \checkmark & 0.143 \\
9 & \path{M:bb_code} & \checkmark & 0.353 & 31 & \path{L:logical_h} & \checkmark & 0.032 \\
10 & \path{M:four_d_geo_code} & \checkmark & 0.062 & 32 & \path{L:logical_h_rotated_init_meas0} & \checkmark & 0.022 \\
11 & \path{M:H_code_family_X} & \checkmark & 0.022 & 33 & \path{L:logical_h_rotated_init_meas1} & \checkmark & 0.022 \\
12 & \path{M:H_code_family_Z} & \checkmark & 0.014 & 34 & \path{L:logical_s_to_the_fourth_is_identity}\textsuperscript{$\dagger$} &  & 0.015 \\
13 & \path{M:H_six_code_X} & \checkmark & 0.021 & 35 & \path{L:state_injection_Y} &  & 0.007 \\
14 & \path{M:H_six_code_Z} & \checkmark & 0.020 & 36 & \path{C:s_gate_teleport_rotated_injection_ZZ} &  & 0.018 \\
15 & \path{M:two_patch_system_has_two_observables} & \checkmark & 0.050 & 37 & \path{C:s_gate_teleport_rotated_injection_cnot_trans} &  & 0.016 \\
16 & \path{M:qec_patch_interface_infers_default_extraction_block} & \checkmark & 0.017 & 38 & \path{L:logical_s_roundtrip} &  & 0.017 \\
17 & \path{M:xzzx_surface_code_X} & \checkmark & 0.019 & 39 & \path{C:s_gate_teleport_unrotated_ZZ_logical_gate} &  & 0.033 \\
18 & \path{M:xzzx_surface_code_Z} & \checkmark & 0.019 & 40 & \path{C:s_gate_teleport_unrotated_cnot_trans_logical_gate} &  & 0.035 \\
19 & \path{L:transversal_cnot} & \checkmark & 0.064 & 41 & \path{C:s_gate_teleport_unrotated_ZZ_inject} &  & 0.027 \\
20 & \path{L:ghz} & \checkmark & 0.139 & 42 & \path{C:s_gate_teleport_unrotated_cnot_trans_inject} &  & 0.025 \\
21 & \path{L:state_injection_X} & \checkmark & 0.022 & 43 & \path{C:ls_distillation_build} &  & 0.567 \\
22 & \path{L:state_injection_Z} & \checkmark & 0.022 & & & & \\
\bottomrule
\end{tabular}

\smallskip
\begin{minipage}{\textwidth}
\footnotesize
$^\dagger$\path{L:logical_s_to_the_fourth_is_identity} is not frame-CSS but
has a common reference when qubits are not reloaded. Its $S$ and CZ gates form nine closed blocks, each of
which multiplies to the identity. A lost atom carries no $X$ component through
its block, so the gates that loss removes would have added no phase, and every
deterministic parity keeps its loss-free value.
\end{minipage}
\end{table}

\begin{table}[!htbp]
\centering
\footnotesize
\setlength{\tabcolsep}{3pt}
\ContinuedFloat
\caption[]{Common-reference certification (continued). Two-round memory circuits generated with qLDPC~\cite{perlin2023qldpc}.}
\begin{tabular}{@{}r p{.27\textwidth} c r @{\hspace{1.5em}} r p{.27\textwidth} c r@{}}
\toprule
No. & Circuit & Frame-CSS & Check & No. & Circuit & Frame-CSS & Check\\
\midrule
44 & \path{steane_7} & \checkmark & 0.017 & 59 & \path{bb_72} & \checkmark & 0.303 \\
45 & \path{qhamming_15} & \checkmark & 0.046 & 60 & \path{gross_144} & \checkmark & 0.663 \\
46 & \path{qreedmuller_1_4} & \checkmark & 0.048 & 61 & \path{two_gross_288} & \checkmark & 1.741 \\
47 & \path{tetrahedral_15} & \checkmark & 0.037 & 62 & \path{mb_14} & \checkmark & 0.039 \\
48 & \path{qgolay_23} & \checkmark & 0.100 & 63 & \path{gala_c3} & \checkmark & 0.025 \\
49 & \path{iceberg_6} & \checkmark & 0.015 & 64 & \path{pair_partition_l3} & \checkmark & 0.012 \\
50 & \path{c4} & \checkmark & 0.010 & 65 & \path{hgp_hamming7} & \checkmark & 0.148 \\
51 & \path{c6} & \checkmark & 0.014 & 66 & \path{chgp_15} & \checkmark & 1.796 \\
52 & \path{manyhypercube_l1} & \checkmark & 0.015 & 67 & \path{cr_15} & \checkmark & 0.704 \\
53 & \path{surface_rot_d3} & \checkmark & 0.018 & 68 & \path{lp_c2} & \checkmark & 0.069 \\
54 & \path{surface_rot_d5} & \checkmark & 0.051 & 69 & \path{bp_hamming7} & \checkmark & 0.156 \\
55 & \path{surface_unrot_d3} & \checkmark & 0.028 & 70 & \path{distbal_steane_x} & \checkmark & 0.045 \\
56 & \path{toric_rot_4} & \checkmark & 0.042 & 71 & \path{qt_c8_rep2} & \checkmark & 0.039 \\
57 & \path{gen_surface_3d_3} & \checkmark & 0.127 & 72 & \path{five_qubit} &  & 0.008 \\
58 & \path{t4_hadamard_96} & \checkmark & 0.431 & 73 & \path{xzzx_surface_d3} & \checkmark & 0.020 \\
\bottomrule
\end{tabular}
\end{table}

\end{document}